\documentclass[12pt]{article}
\def\vc{\par}
\def\vu{\vskip1.cm}
\def\be{\begin{equation}}
\def\ee{\end{equation}}
\begin{document}
\begin{center}
\hfill \vbox{
\hbox{DFTT 2/2001}
\hbox{DFCAL-TH 01/1}
\hbox{February 2001}}
\vskip 0.5cm
{\Large\bf 
Finite-size scaling and the deconfinement transition in  gauge theories.}\\ 
\end{center}
\vskip 0.6cm
\centerline{R. Fiore$^{a~\dagger}$,  
A. Papa$^{a~\dagger}$ and P. Provero$^{b,c~\ddagger}$}
\vskip 0.6cm
\centerline{\sl $^a$ Dipartimento di Fisica, Universit\`a della Calabria}
\centerline{\sl Istituto Nazionale di Fisica Nucleare, Gruppo collegato
di Cosenza}
\centerline{\sl I--87036 Arcavacata di Rende, Cosenza, Italy}
\vskip0.2cm
\centerline{\sl $^b$ Dipartimento di Scienze e Tecnologie Avanzate}
\centerline{\sl Universit\`a del Piemonte Orientale}
\centerline{\sl I-15100 Alessandria, Italy}
\vskip0.2cm
\centerline{\sl $^c$ Dipartimento di Fisica
Teorica dell'Universit\`a di Torino}
\centerline{\sl Istituto Nazionale di Fisica Nucleare, Sezione di Torino}
\centerline{\sl via P. Giuria 1, I--10125 Torino, Italy}
\vskip 0.6cm
\begin{abstract}
We introduce a new method for determining the critical indices of the
deconfinement transition in gauge theories. The method is based on
the finite size scaling behavior of the expectation value of simple 
lattice operators, such as the plaquette.
We test the method for the case of $SU(3)$ pure gauge theory in 
$(2+1)$ dimensions and obtain a precise determination of the critical index 
$\nu$, in agreement with the prediction of the Svetitsky-Yaffe conjecture.
\end{abstract}
\vfill
\hrule
\vskip.3cm
\noindent
$^{\ast}${\it Work supported by the Ministero italiano
dell'Universit\`a  e della Ricerca Scientifica e Tecnologica.}
\vfill

\noindent$^\dagger$ {\it e-mail address}: fiore,papa@cs.infn.it \newline
$^\ddagger$ {\it e-mail address}: provero@to.infn.it

\newpage

\noindent
Gauge theories at finite temperature undergo a phase transition 
from a low temperature phase, in which color sources are confined by a linearly
rising potential, to a high temperature phase in which the attractive
force is screened at long distances and isolated quarks can exist.
This same behavior can be observed even in pure gauge theories, without
matter fields, by studying the theory in presence of static sources.  
\vc
The order parameter of the deconfinement transition in a pure gauge
theory is the Polyakov loop~\cite{Polyakov:1978vu,Susskind:1979up}, 
representing the world-line of a static source of color charge (that can be 
thought of as an infinitely massive quark). The Polyakov loop takes 
values in the center of the gauge group: the $(d+1)$-dimensional gauge 
theory generates an effective $d$-dimensional theory for the Polyakov loop, 
having the center of the gauge group as global symmetry group.
The deconfinement phase transition coincides with the spontaneous breaking of 
this global symmetry.
\vu
In this work we introduce a new method for the computation of the
critical indices of a pure gauge theory with second order deconfinement 
transition. According to the Svetitsky-Yaffe  
conjecture~\cite{Svetitsky:1982gs},
such critical indices for the $(d+1)$-dimensional gauge theory 
coincide with those of the effective $d$-dimensional model, if the latter
has also a second order phase transition. We will consider in particular 
the critical index $\nu$ of the correlation length.
For a general $d$-dimensional statistical model a possible way of
extracting the value of $\nu$ from lattice Monte Carlo simulations is to study
the finite size scaling (FSS) behavior of the energy operator: 
FSS theory predicts that, if $L$ is the lattice size, the expectation value
of the energy operator behaves for large $L$ as
\be
\langle E\rangle_L\sim \langle E\rangle_\infty+k L^{\frac{1}{\nu}-d}\;,
\label{efss}
\ee
where $\langle E\rangle_\infty$ is the  expectation value of the
energy operator in the thermodynamic limit and $k$ is a non-universal
constant. This method was applied {\it e.g.} in Ref.~\cite{Hasenbusch:1997} to
evaluate the critical index $\nu$ of the three-dimensional Ising
model.
\vc
Compared to other methods based on the FSS of fluctuation operators
such as susceptibilities or Binder cumulants, the advantage lies in
the fact that $\langle E\rangle_L$ can be computed to high
accuracy. The main drawback is that the term containing $\nu$ in
Eq.(\ref{efss}) is subdominant for $L\to\infty$ with respect to the
bulk contribution $\langle E\rangle_\infty$. The numerical results we
will present strongly suggest that, in the case of gauge theories, the
advantages outweigh the drawbacks and the method can give very accurate
results.  
\vu
Consider a gauge invariant operator  $\hat{O}$
that is invariant also under the 
global symmetry given by the center of the gauge group. 
Examples of such operators include Wilson loops and those 
of the form
\be
P(x) P^{\dagger}(y)\;,
\label{pp}
\ee
where $P(x)$ and $P(y)$ are Polyakov loops at two different sites
$x$ and $y$ of the $d$-dimensional space. It is natural to expect
the operator product expansion (OPE) of any such operator to have the same
form as the one of $E$:
\be
\hat{O}=c_I\ I + c_\epsilon\  \epsilon 
+\cdots \;.
\label{ope}
\ee
Here $I$ and $\epsilon$ are, respectively, the identity and the {\em
scaling} energy operator\footnote{Not to be confused with the {\em
lattice} energy operator $E$, which is just an example of an operator
with OPE given by Eq.~(\ref{ope}).} 
in the statistical model, and the dots represent contributions of 
irrelevant operators. 
The ansatz of Eq.~(\ref{ope}) was introduced and tested in 
Ref.~\cite{Gliozzi:1997yc}, and used in Ref.~\cite{Fiore:1998uk,Fiore:1999fs},
to obtain 
some exact results on correlation functions of $(2+1)$-dimensional 
gauge theories at the deconfinement transition.
\vc 
In particular, Eq.~(\ref{ope}) implies that the FSS behavior 
of the expectation value $\langle \hat{O} \rangle$ will have the form of 
Eq.~(\ref{efss}):
\be
\langle \hat{O}\rangle_L\sim \langle \hat{O}\rangle_\infty
+c L^{\frac{1}{\nu}-d}\;.
\label{ofss}
\ee
The contributions of the irrelevant operators will be subleading for 
$L\to \infty$.
\vc
We conclude that the FSS behavior of any such operator
can be used to determine the value of $\nu$ through
Eq.~(\ref{ofss}). The obvious advantage is that one can use operators,
such as the plaquette, whose expectation value can be computed to high
accuracy with relatively modest computational effort. 
\vu
To test the method in practice, we used it to determine the
critical index $\nu$ for $SU(3)$ pure gauge theory in $(2+1)$-dimensions. 
This was evaluated from Monte Carlo simulations in Ref.~\cite{Engels:1997dz} 
with the result
\be
\nu_{MC}=0.90(20)\;.
\ee
However, if 
the Svetitsky-Yaffe conjecture~\cite{Svetitsky:1982gs} holds then 
the 2-dimensional effective theory of the Polyakov loop 
is in the universality class of the 3-state Potts model, and
the value of $\nu$ is known exactly:
\be
\nu=\frac{5}{6}=0.833\cdots\label{sy}\;.
\ee
\vc
The observables we considered are the 
expectation values of the plaquette and of the operator defined in
Eq.~(\ref{pp}),
where $x$ and $y$ are taken to be nearest 
neighbors sites in the 2-dimensional (spatial) lattice. We computed these 
expectation values on lattices with temporal extension 
$N_t=2$ and spatial sizes 
$L=N_x=N_y$ ranging from $7$ to $30$. The simulations were performed at the
critical coupling
\be
\beta_c(N_t=2)=8.155
\ee   
obtained in Ref.~\cite{Engels:1997dz}. The simulation algorithm we adopted
was a mixture of one sweep of a 10-hit Metropolis and four sweeps of
over-relaxation consisting in two updates of (random) $SU(2)$ subgroups.
For each simulation we collected 400K equilibrium configurations, separated
each other by 10 updating steps. The error analysis was performed by the
jackknife method applied to data bins at different levels of blocking.
\vc
We report in Table~1 the expectation values we obtained. Because of the
asymmetry of the lattice, space-like and time-like plaquettes have
obviously different expectation values and must be considered as two
different operators.
To evaluate $\nu$, we performed a single multibranched fit of the
three data sets. To avoid cross-correlations we included
in the fit only space-like (``magnetic'') plaquettes from lattices
with odd $L$ and time-like (``electric'') plaquettes from lattices 
with even $L$. Polyakov loop correlations were measured in separate runs and 
therefore are not correlated with the plaquette measurements.
\vc
The result of the fit is
\be
\nu=0.827(22)\;, \qquad\qquad\chi^2_{\rm red}=0.84\;,
\ee
in excellent agreement with the value of Eq.~(\ref{sy}), coming
from the Svetitsky-Yaffe conjecture, and with remarkably improved accuracy 
in comparison to the existing Monte Carlo evaluation. 
\vu
In conclusion, the method we have proposed appears to give very
precise results for critical indices, thanks to the fact that only
``simple'' expectation values have to be evaluated, such as the plaquette 
expectation value.
The same method can be used to evaluate the critical
coupling (that, in our case, was taken from the literature): 
since Eq.~(\ref{ofss}) is valid only at the critical point, one should
perform the same fit at several couplings and determine the value of the 
critical one, by comparing the $\chi^2$'s of the different fits.
\vu\noindent{\bf Acknowledgements.} We are grateful to F. Gliozzi for
many enlightening discussions. One of us (P.P.) thanks M. Caselle
for helpful and stimulating conversations. We also thank J. Engels for
comments on a previous version of this work.

\begin{table}[ht]
\caption{\sl
Expectation values of the electric and magnetic plaquette and
Polyakov loop correlator.}
\label{mcrplaq2}
\begin{center}
\begin{tabular}{|c|c|c|c|}
\hline
$L$ &Plaquette (el.)  & Plaquette (mag.)&$PP$ correl.\\
\hline
 7&0.628855(23)&0.625721(22)&\\
 8&0.628562(30)&0.625494(27)&1.2707(13)\\
 9&0.628318(26)&0.625358(24)&1.2530(15)\\
10&0.628107(27)&0.625180(25)&1.2358(15)\\
11&0.627929(26)&0.625059(21)&\\
12&0.627825(25)&0.624983(21)&1.2122(14)\\
13&0.627729(26)&0.624893(22)&1.1997(15)\\
14&0.627593(23)&0.624821(20)&1.1944(16)\\
15&0.627493(24)&0.624741(19)&\\
16&0.627416(22)&0.624692(19)&1.1829(15)\\
17&0.627343(23)&0.624643(19)&\\
18&0.627270(22)&0.624598(17)&1.1669(16)\\
19&0.627212(23)&0.624545(18)&1.1625(16)\\
20&0.627182(22)&0.624523(17)&1.1573(16)\\
21&0.627121(23)&0.624482(19)&\\
22&0.627067(20)&0.624445(16)&1.1511(15)\\
23&0.627007(23)&0.624396(17)&\\
24&0.627043(23)&0.624426(17)&1.1429(16)\\
25&0.626977(23)&0.624385(17)&1.1408(17)\\
26&0.626922(22)&0.624338(16)&1.1390(16)\\
28&0.626855(24)&0.624297(17)&1.1322(16)\\
29&0.626869(22)&0.624311(17)&1.1311(19)\\
30&0.626814(26)&0.624265(18)&1.1289(17)\\
\hline

\end{tabular}
\end{center}
\end{table}
\end{document}